\documentclass[apjl,twocolappendix]{emulateapj}
\usepackage{graphicx,color}
\usepackage{amssymb,txfonts}
\usepackage{epsfig,natbib}
\newcommand{\white}{\color{white}}
\newcommand{\asr}{{Adv. Space Res.}}
\newcommand{\lrsp}{{Liv. Rev. Sol. Phys.}}
\DeclareMathAlphabet{\mathitbf}{OML}{cmm}{b}{it}
\DeclareMathAlphabet{\mathf}{OML}{cmm}{c}{sl}

\newcommand{\Bz}{{\mathf B_z}}
\newcommand{\Bzabs}{\mathit B_z}
\newcommand{\Bhabs}{\mathit B_h}
\newcommand{\Bhave}{\overline{\Bhabs}}
\newcommand{\BhaveHMI}{{\overline{\mathit B_h}}_{\rm, HMI}}
\newcommand{\BhaveSPbin}{{\overline{\mathit B_h}}_{\rm, SP_{\rm bin}}}
\newcommand{\BhaveSPorig}{{\overline{\mathit B_h}}_{\rm, SP_{\rm orig}}}
\newcommand{\Brave}{<\hspace{-2pt}\Bzabs/\Bhabs\hspace{-2pt}>}
\newcommand{\dBhave}{\Delta\Bhave}
\newcommand{\Enlff}{E_{\rm nlff}}
\newcommand{\Epot}{E_{\rm pot}}
\newcommand{\Eexcess}{\Delta E}

\newcommand{\fplb}{\widetilde{F}_{\rm p}}

\newcommand{\fuv}{|\phi_z|}
\newcommand{\fuvp}{|\phi'_z|}
\newcommand{\lfl}{l_{\rm FL}}
\newcommand{\hmax}{h_{\rm max}}

\newcommand{\da}{~{\rm d}\mathcal A}
\newcommand{\km}{{\rm\,km}}

\newcommand{\mT}{{\rm\,mT}}

\newcommand{\Mx}{{\rm\,Mx}}
\newcommand{\J}{{\rm\,J}}

\newcommand{\kgm}{{\rm\,kg\,m^{-3}}}

\newcommand{\Mm}{{\rm\,Mm}}
\newcommand{\degree}{^\circ}

\newcommand{\ie}{i.\,e.}
\newcommand{\eg}{e.\,g.}
\newcommand{\etal}{et al.}
\slugcomment{Accepted for publication in ApJ}
\shorttitle{Magnetic field modeling using HMI and SP data}
\shortauthors{Thalmann \etal}

\begin{document}

\title{Comparison of force-free coronal magnetic field modeling\\
using vector fields from Hinode and Solar Dynamics Observatory}

\author{J.~K.~Thalmann\altaffilmark{1}, S.~K.~Tiwari\altaffilmark{1} and T.~Wiegelmann\altaffilmark{1}}
\email{thalmann@mps.mpg.de}
\altaffiltext{1}{Max-Plank-Institut f\"ur Sonnensystemforschung, Max-Planck-Str. 2, 37191 Katlenburg-Lindau, Germany}

\begin{abstract}
Photospheric magnetic vector maps from two different instruments are used to model the nonlinear force-free coronal magnetic field above an active region. We use vector maps inferred from polarization measurements of the Solar Dynamics Observatory/Helioseismic and Magnetic Imager (HMI) and the Solar Optical Telescope Spectropolarimeter (SP) aboard Hinode. Besides basing our model calculations on HMI data, we use both, SP data of original resolution and scaled down to the resolution of HMI. This allows us to compare the model results based on data from different instruments and to investigate how a binning of high-resolution data effects the model outcome. The resulting 3D magnetic fields are compared in terms of magnetic energy content and magnetic topology. We find stronger magnetic fields in the SP data, translating into a higher total magnetic energy of the SP models. The net Lorentz forces of the HMI and SP lower boundaries verify their force-free compatibility. We find substantial differences in the absolute estimates of the magnetic field energy but similar relative estimates, e.g., the fraction of excess energy and of the flux shared by distinct areas. The location and extension of neighboring connectivity domains differs and the SP model fields tend to be higher and more vertical. Hence, conclusions about the magnetic connectivity based on force-free field models are to be drawn with caution. We find that the deviations of the model solution when based on the lower-resolution SP data are small compared to the differences of the solutions based on data from different instruments.
\end{abstract}


\section{Introduction}

Most recent studies dealing with the magnetic structure of the solar corona above active regions use different force-free model approaches \citep[see recent review of][]{wie_sak_12} and base the modeling on photospheric vector magnetic field data from either the Helioseismic and Magnetic Imager (HMI) on board the Solar Dynamics Observatory (SDO) or the Spectropolarimeter (SP) of the Solar Optical Telescope (SOT) on board the Hinode spacecraft. Such model approaches are used to compensate the lack of routine direct measurements of the coronal magnetic field vector.

\cite{hao_guo_12} used SP data as lower boundary condition to an optimization approach to analyze the coronal magnetic field associated to a white light flare. The modeling suggested that the flare originated from sheared and twisted field lines with low altitudes bridged by a set of higher magnetic field lines. \cite{ino_shi_12} applied a MHD relaxation method based on SP data to investigate the buildup and release of magnetic twist and suspected the importance of the relative handedness of twisted field lines and the ambient field. 

\cite{sun_hoe_12a} employed an optimization method based on HMI data to model the temporal evolution of the coronal field of an active region over five days. The modeling displayed distinct stages of the build-up and release of magnetic energy and analyzed the association to changes in the magnetic field. In a subsequent study, \cite{sun_hoe_12b} used the same method for a detailed analysis of the field topology during a series of eruptions observed from HMI. Above the apexes of cusp-like loops observed in coronal images, the modeling result suggested the presence of a coronal null point.

Comparisons of the outcome of {\it different} force-free model algorithms based on the {\it same} lower boundary conditions have been performed in the past too \citep[\eg,][]{der_schr_09,gil_whe_12}. These studies revealed that order-of-magnitude estimates of these models based on large enough fields-of-view and high enough spatial resolution of the vector magnetic field data can be expected to be reliable.

The model outcome of the {\it same} reconstruction algorithm based on data from {\it different} instruments has been studied recently by \cite{tha_pie_12}, using data from HMI and the Vector-SpectroMagnetograph (VSM) of the SOLIS project \citep{kel_har_03a}. They found agreements of the resulting force-free models in form of, \eg, the relative amount of energy to be set free during an eruption but also found a considerable difference in the absolute model energy estimates. In particular, they found the estimated energy content of the VSM model being about twice of that of the HMI model.

In the present study we use vector magnetic field data of HMI and SP as an input to the same force-free model and compare the results. This is motivated due to the data of these two instruments being widely used recently and in the future expected to be frequently used as an input for the modeling of the coronal magnetic field. We regard this as important in order to test the consistency of the model solutions and at the same time to give a feeling about the accuracy of the estimated physical quantities and magnetic field topology based on those models. We also investigate the effect which a binning of the SP instrument data to a lower resolution has on the model outcome. We, however, do not search for explanations of the differences of the HMI and SP inversion products itself, considering this as to be out of the scope of this study.

\section{Data and Analysis methods}

\subsection{Data Sources}

The HMI on board SDO \citep{scho_sche_12} obtains filtergrams at the photospheric Fe\,{\sc{i}} 617.3\,nm spectral line. The full Stokes vector is retrieved from filtergrams averaged over about 12 min and inverted using the Milne-Eddington (ME) inversion algorithm of \citet{bor_tom_10}. The 180$\degree$-azimuth ambiguity of the transverse field is resolved using a Minimum Energy Algorithm \citep{met_94, met_lek_06, lek_bar_09} and the resulting vector magnetograms have a plate scale of 0.5~arc-second.

In its fast scan mode, SP as part of the SOT \citep{tsu_08,sue_tsu_08,ich_lit_08,shi_nag_08} on board the Hinode spacecraft \citep{kos_mat_07} obeys a spatial resolution of $\sim$\,0.32~arc-second. It observes the Fe\,{\sc{i}} spectral line couplet at 630.15\,nm and 630.25\,nm. Full Stokes profiles are obtained with a spectral sampling of 2.25\,$10^{-3}$\,nm and a slit scan time of 1.6\,s. The physical parameters from the full Stokes profiles were obtained using the MERLIN ME inversion algorithm \citep{sku_lit_87,lit_cas_07}. The 180$\degree$-azimuth ambiguity is resolved in the same way as for HMI data.

\subsection{Event Selection and Data Set}\label{ss:event_selection_and_data_set}

To perform a study as described above simultaneous observations from HMI and SP are required. Both, HMI and SP vector magnetic field data are available to analyze the magnetic structure of active region~11382 on 2011 December 22. SP scanned this active region from 04:46~UT to 05:29~UT (\ie, scanned $\sim$\,$3\Mm$/min from solar east to west). The HMI vector map used for this study was retrieved at $\tau_{\rm rec}$\,$=$\,05:00~UT, approximately at half of the scanning time of SP. Given average photospheric conditions ($B$\,$\sim$\,$100\mT$, $\rho$\,$\sim$\,$10^{-4}$\,$\kgm$) the Alfv{\'e}n speed is $v_A$\,$\sim$\,10\,km\,s$^{-1}$. The Alfv{\'e}n travel time and thus the evolution of the photospheric magnetic field over a characteristic distance in the present study ($\sim$\,$100\Mm$) is $\sim$\,1\,h. Therefore, we assume that the temporal changes of the photospheric field over $\sim$\,20~min, during which SP scanned before and after the time when HMI recorded, are negligible. Regarding the longitudinal magnetic field we assume this as justified since, \eg, also \cite{wan_zha_09a} found that the average ratio of the longitudinal magnetic flux densities measured by SP and SoHO/MDI \citep{sche_95} was not strongly influenced by the evolution of the photospheric magnetic field during the scanning time of SP.

The magnetic field vectors are transformed to Heliographic coordinates \citep{gar_hag_90}. We correct the SP data set for the effect of differential rotation where we use $\tau_{\rm rec}$ as reference time. Given present computational capabilities high-resolution data is sometimes binned to a lower resolution in order to allow for near real-time magnetic field modeling and/or computational domains of feasible dimensions \citep[see \eg,][]{der_schr_09}. Thus, in the course of co-alignment of the HMI and SP data, we also bin the original-resolution SP (SP$_{\rm orig}$) data to the resolution of HMI which involves a 2D linear interpolation. The binned SP data is hereafter denoted as SP$_{\rm bin}$ data.

\begin{figure}
	\centering
	\includegraphics[width=\columnwidth]{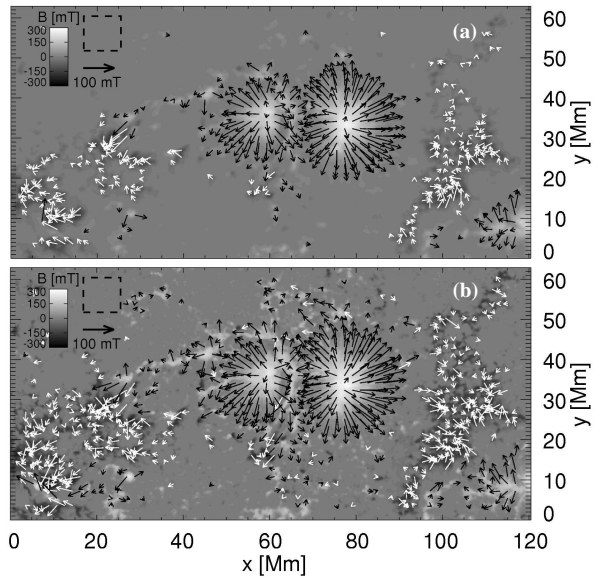}
	\put(-55,205){\bf\white(a)}
	\put(-55,107){\bf\white(b)}
	\caption{(a) HMI and (b) SP$_{\rm bin}$ data. The gray-scale background reflects the vertical magnetic field, $\Bz$. Black/white arrows indicate the direction of the horizontal magnetic field, originating from positive/negative polarity where $\Bzabs$\,$\ge$\,$20\mT$. The dashed rectangle outlines the quiet-Sun region used to calculate the 2$\sigma$ uncertainty of the vertical and horizontal magnetic field. The black arrow just below the dashed rectangle indicates the length of an arrow representing a horizontal field magnitude of $100\mT$.}
	\label{fig:fig1}
\end{figure}

The field-of-view (FOV) used to study active region~11382 is mainly determined by the area covered by the SP scan ($\sim$\,140\,$\times$\,85$\Mm^2$ centered around S19W07; see Figure~\ref{fig:fig1}). Within the HMI data which covers roughly $\sim$\,400\,$\times$\,250$\Mm^2$, we define a window equally sized as the FOV of SP and calculate the cross-correlation between its vertical field component and that of the SP data. By shifting the position of the HMI window we search for the highest cross-correlation to find the corresponding HMI sub-field.

\subsection{Uncertainty Estimation}\label{ss:uncertainty_estimation}

The noise level of the HMI data is on the order of $1\mT$/$10\mT$ for the longitudinal/transverse magnetic field (X.~Sun, private communication). The average uncertainty for both the longitudinal and transverse field, estimated from the SP inversion error maps is $\simeq$\,$1\mT$. Besides seeming rather low especially for the transverse field these standard error estimates may not be reliable (B.~Lites, private communication). Thus, we employ a consistent measure of the uncertainty level for the data from the two instruments, as described in the following.

When investigating the properties of the HMI and SP data in the course of the force-free modeling, we only consider ({\sc{i}}) pixels with values of the vertical field, $\Bzabs$, {\it and} horizontal field, $\Bhabs$, above the respective 2$\sigma$ uncertainty levels (2$\sigma_{\Bzabs}$ and 2$\sigma_{\Bhabs}$, respectively) or ({\sc{ii}}) pixels with values of $\Bzabs$\,$>$\,2$\sigma_{\Bzabs}$ {\it and} $\Bhabs$\,$<$\,2$\sigma_{\Bhabs}$, the latter in order not to disregard the strong vertical fields in the center of the active region. 

The respective 2$\sigma$ uncertainty levels are calculated for the quiet-Sun area, outlined by the dashed rectangles in Figure~\ref{fig:fig1}, where we find: 2$\sigma_{\Bzabs}$\,$=$\,3.3/6.7$\mT$ and 2$\sigma_{\Bhabs}$\,$=$\,5.2/12.2$\mT$ for the HMI/SP$_{\rm bin}$ data. The uncertainty levels for the SP$_{\rm orig}$ data are calculated from a quiet-Sun region, equivalent to that outlined in Figure~\ref{fig:fig1}. Here we find 2$\sigma_{\Bzabs}$\,$=$\,8.8$\mT$ and 2$\sigma_{\Bhabs}$\,$=$\,12.3$\mT$, \ie~slightly higher than what was found for the SP$_{\rm bin}$ data. The latter estimates, in fact, conform with the findings of \cite{oro_bel_07} who had estimated the measurement uncertainties of SP for the quiet-Sun inter network field strengths and fluxes as $<$\,15$\mT$.

\begin{figure*}
	\epsscale{0.65} 
	\plotone{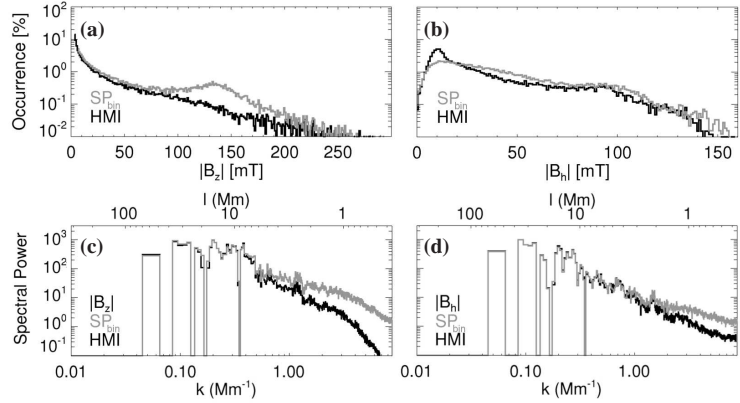}
	\put(-255,140){\bf(a)}
	\put(-125,140){\bf(b)}
	\put(-255,58){\bf(c)}
	\put(-125,58){\bf(d)}
	\caption{Relative occurrence of the absolute (a) vertical field, $\Bzabs$, and (b) horizontal field, $\Bhabs$. The power of $\Bzabs$ and $\Bhabs$ as a function of length scale, $l$, and wave number, $k$, is shown in panels (c) and (d), respectively. HMI/SP$_{\rm bin}$ data are represented by black/gray lines. Only pixels with values $\Bzabs$\,$>$\,2$\sigma_{\Bzabs}$ {\it and} $\Bhabs$,$>$\,2$\sigma_{\Bhabs}$ or $\Bzabs$\,$>$\,2$\sigma_{\Bzabs}$ {\it and} $\Bhabs$\,$<$\,2$\sigma_{\Bhabs}$ are taken into account.}
	\label{fig:fig2}
\end{figure*}

\subsection{Magnetic Field Modeling}

Photospheric polarization signals originate from atmospheric layers which are known not to be force-free. For instance, \cite{met_95} showed, using vector magnetic field measurements of an active-region magnetic field, that it can be considered as to be force-free above 0.4~Mm above a photospheric level. Modeling the interchanging dominance of plasma and magnetic pressure, \cite{gar_01} was able to estimate the height regime of dominating magnetic fields above a sunspot/plage region as $\sim$\,0.8 -- 200~Mm. Using high-resolution SP vector maps, \cite{tiw_12} found that umbral and inner penumbral parts of sunspots may nearly be force-free but that sunspots as a whole at a photospheric level might not entirely be so.

Therefore, the inferred HMI and SP magnetic vector maps are not force-free consistent and need to be preprocessed to achieve suitable force-free consistent boundary conditions \citep{wie_inh_06}. From the vertical component of the preprocessed field vector, a potential field is calculated and used as start equilibrium and to prescribe the boundaries of the cubic computational domain. The bottom boundary is replaced by the preprocessed vector field and the set of force-free equations for the nonlinear case in Cartesian coordinates solved \citep[][]{wie_inh_10}. A boundary layer of $\sim$\,$10\Mm$ is introduced towards the lateral and top boundaries where the nonlinear force-free (NLFF) solution drops to the prescribed boundary field. For our analysis we discard this layer and only consider the inner (physical) $\sim$\,120\,$\times$\,60\,$\times$\,70$\Mm^3$ domain and the according bottom boundary field. Since this method involves the relaxation of the magnetic field not only inside the computational domain but also on its bottom boundary we compute a potential field from the relaxed lower boundary based on the Fast-Fourier method described by \cite{ali_81}.

Hereafter, we refer to the 3D NLFF field model based on the HMI vector map as an input to as ``HMI model''. Similarly, the ``SP$_{\rm bin}$ model'' and ``SP$_{\rm orig}$ model'' result from using the binned and original-resolution SP vector magnetic field, respectively, as input data to our preprocessing and force-free reconstruction algorithms. We compare the HMI and SP$_{\rm bin}$ model in \S\,\ref{ss:mfaff}--\ref{ss:topo} and summarize the effects of binning on the model outcome in \S\,\ref{ss:eof_binning}.

\section{Results}

\subsection{Magnetic Flux and Force-Freeness}\label{ss:mfaff}

After performing the data preparation as described in \S~\ref{ss:event_selection_and_data_set}, we are able to investigate how the vertical and horizontal field components of the HMI and SP$_{\rm bin}$ vector maps compare to each other and to check the force-free consistency. We only consider data points where the criteria outlined in \S~\ref{ss:uncertainty_estimation} are fulfilled.

We find on overall stronger vertical fields in the SP$_{\rm bin}$ than the HMI data (especially for $\Bzabs$\,$\gtrsim$\,$100\mT$; see Figure~\ref{fig:fig2}a) and stronger horizontal field (except for $\Bhabs$\,$\lesssim$\,$10\mT$; see Figure~\ref{fig:fig2}b). This can be seen also when comparing the area-integrated unsigned vertical flux, $\fuv$, and the average horizontal field, $\Bhave$: the HMI data hosts $\sim$\,54\% of $\fuv_{\rm SP_{\rm bin}}$ and $\sim$\,81\% of $\BhaveSPbin$ (see Table~\ref{tab:forces}). HMI and SP$_{\rm bin}$ have a comparable sensitivity on long scales, $l$, \ie~at small wave numbers, $k$, especially on scales $l$\,$\gtrsim$\,$5\Mm$ (Figure~\ref{fig:fig2}c,d). With decreasing scale the amount of detected field increasingly differs: SP data show considerably stronger fields on smaller scales.

\begin{figure*}
	\epsscale{0.65} 
	\plotone{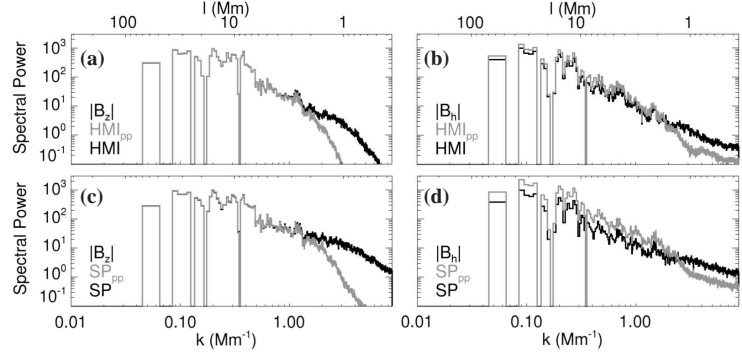}
	\put(-255,110){\bf(a)}
	\put(-125,110){\bf(b)}
	\put(-255,58){\bf(c)}
	\put(-125,58){\bf(d)}
	\caption{Power of the absolute vertical field, $\Bzabs$, and horizontal field, $\Bhabs$, before and after applying a preprocessing to it (black and gray curves, respectively). Panels (a)/(c) show the power of $\Bzabs$ of the HMI/SP$_{\rm bin}$ data while panels (b)/(d) show that of $\Bhabs$ as a function of length scale, $l$, and wave number, $k$. Only pixels with values $\Bzabs$\,$>$\,2$\sigma_{\Bzabs}$ {\it and} $\Bhabs$,$>$\,2$\sigma_{\Bhabs}$ or $\Bzabs$\,$>$\,2$\sigma_{\Bzabs}$ {\it and} $\Bhabs$\,$<$\,2$\sigma_{\Bhabs}$ are taken into account.}
	\label{fig:fig3}
\end{figure*}

\begin{table}
	\caption{\label{tab:forces} Magnetic field related quantities of the vector data}
	\vspace{-11pt}
	\begin{center}
		\begin{tabular}{lccr@{.}lr@{.}lr@{.}l}
			\tableline
			~ & $\fuv$ & $\Bhave$ & \multicolumn{2}{c}{$\frac{|\widetilde{F}_x|}{|\fplb|}$} & \multicolumn{2}{c}{$\frac{|\widetilde{F}_y|}{|\fplb|}$} & \multicolumn{2}{c}{$\frac{|\widetilde{F}_z|}{|\fplb|}$} \\
			~ & $[\,\times$\,$10^{22}$\,$\Mx\,]$ & $[\,\mT\,]$ & \multicolumn{6}{c}{~}\\
			\tableline
			HMI & 1.323 & 18.4 & 0&13 & 0&08 & 0&29 \\
			SP$_{\rm bin}$ & 2.440 & 22.7 & 0&03 & 0&06 & 0&81 \\
			SP$_{\rm orig}$ & 2.510 & 21.5 & 0&03 & 0&06 & 0&84 \\
			\multicolumn{9}{c}{$Preprocessed~data$}\\
			HMI & 1.280 & 19.2 & 0&01 & 0&01 & 0&03 \\
			SP$_{\rm bin}$ & 2.341 & 30.2 & 0&02 & 0&01 & 0&02\\
			SP$_{\rm orig}$ & 2.427 & 30.1 & 0&02 & 0&02 & 0&02 \\
			\multicolumn{9}{c}{$NLFF~lower~boundary~data$}\\
			HMI & 1.279 & 20.9 & 0&01 & 0&02 & 0&01 \\
			SP$_{\rm bin}$ & 2.338 & 33.5 & 0&01 & 0&03 & 0&01 \\
			SP$_{\rm orig}$ & 2.364 & 30.3 & 0&01 & 0&03 & 0&08 \\
			\tableline
		\end{tabular}
	\end{center}
	{Listed are the total unsigned vertical flux, $\fuv$, the average horizontal field, $\Bhave$ as well as the net Lorentz force components ($|\widetilde{F}_x|$, $|\widetilde{F}_y|$ and $|\widetilde{F}_z|$) normalized to the magnetic pressure force, $|\fplb|$. Only pixels with values $\Bzabs$\,$>$\,2$\sigma_{\Bzabs}$ {\it and} $\Bhabs$,$>$\,2$\sigma_{\Bhabs}$ or $\Bzabs$\,$>$\,2$\sigma_{\Bzabs}$ {\it and} $\Bhabs$\,$<$\,2$\sigma_{\Bhabs}$ are taken into account.}
\end{table}

A necessary condition for a magnetic field to be force-free is that the components of the net Lorentz force are considerably smaller than a characteristic magnitude of the total Lorentz force in case of a non force-free magnetic field. The latter can be approximated by the magnetic pressure, $\fplb$, on the lower boundary \citep{low_84}. The ratio $|\widetilde{F}_i|/|\fplb|$ with $i$\,$=$\,$(x,y,z)$ in Table~\ref{tab:forces} shows that this conditions are met only to a certain degree. The ratios found here agree with the values found by, \eg, \cite{moo_02} and \cite{tiw_12} and we also find $|\widetilde{F}_z|$\,$>$\,[\,$|\widetilde{F}_x|$, $|\widetilde{F}_y|$\,]. Preprocessing, however, certainly improves the force-freeness as it smooths the vertical field and, additionally, alters the horizontal field in order to minimize the net force and torque and to gain boundary conditions compatible with the force-free assumption (showing ratios of clearly smaller than unity). The effect of smoothing can be seen when comparing the spectral power of the raw and preprocessed HMI and SP$_{\rm bin}$ data (Figure~\ref{fig:fig3}): the signal on shorter scales (\ie, for large $k$) is reduced. On longer scales, the power of the vertical field remains the same (see Figure~\ref{fig:fig3}a,c) but that of the horizontal field is enhanced (more pronounced for the SP$_{\rm bin}$ data; see Figure~\ref{fig:fig3}b,d). 

The preprocessing leads to a slight reduction of $\fuv$: $\fuv_{\rm HMI}$ is reduced by $\sim$\,3\% and $\fuv_{\rm SP}$ by $\sim$\,4\%. The preprocessing also leads to an enhancement of $\Bhave$: $\BhaveHMI$ is enhanced by $\sim$\,4\% and $\BhaveSPbin$ is enhanced by $\sim$\,30\% (see Table~\ref{tab:forces}). The preprocessed HMI data hosts $\sim$\,55\% of $\fuv_{\rm SP}$ and $\sim$\,64\% of $\BhaveSPbin$ of the preprocessed SP$_{\rm bin}$ data. In summary, the preprocessing only slightly reduces the difference of the vertical unsigned flux between HMI and SP$_{\rm bin}$ (which is $\sim$\,45\% before as well as after preprocessing) but enhances the difference of the average horizontal field, $\dBhave$, (from $\dBhave$\,$\sim$\,19\% {\it before} to $\dBhave$\,$\sim$\,36\% {\it after} preprocessing). These preprocessed HMI and SP$_{\rm bin}$ data are used as lower boundary condition for the NLFF reconstruction.

As mentioned above, our NLFF modeling algorithm also relaxes the magnetic field on the bottom boundary of a cubic computational domain. Thus, besides during the preprocessing, the magnetic field vector on the lower boundary is altered also while iteratively seeking for the force- and divergence-free field solution in the volume. It's modification to the preprocessed lower boundary data, as listed in Table~\ref{tab:forces}, is that $\fuv$ is reduced by $\lesssim$\,0.1\% and $\Bhave$ is increased by $\sim$\,10\%. Thus, the HMI NLFF lower boundary hosts $\sim$\,55\% of $\fuv_{\rm SP}$ and $\sim$\,62\% of $\BhaveSPbin$ of the SP$_{\rm bin}$ NLFF lower boundary data, comparable to the ratio we found for the preprocessed data.

\subsection{Magnetic Energy}\label{ss:merg}

From the 3D model fields, we can estimate the magnetic energy content of the potential field ($\Epot$) and of the NLFF field (total energy; $\Enlff$). An upper limit for the energy which can be released (excess energy) is given by $\Eexcess$\,$=$\,$\Enlff$\,$-$\,$\Epot$. We can estimate the statistical accuracy of our volume-integrated energy estimates by adding different artificial noise models to the HMI and SP magnetograms, consecutive application of the preprocessing and extrapolation algorithms and comparison of the resulting energy values. This yields a statistical error of $\sim$\,1\% for both $\Epot$ and $\Enlff$ and $\sim$\,10\% for $\Eexcess$.

The estimated {\it absolute} potential energy of the HMI model is $\sim$\,48\% of that of the SP$_{\rm bin}$ model (see Table~\ref{tab:energies}). This is a direct consequence of the HMI NLFF bottom boundary hosting only $\sim$\,55\% of the unsigned vertical flux of the SP$_{\rm bin}$ bottom boundary (see Table~\ref{tab:forces}). A similar trend is found for the {\it absolute} total and excess energy of the HMI model. 

However, the {\it relative} excess energy is about 20\% of the total energy in both models (given by the ratio $\Eexcess/\Enlff$ in Table~\ref{tab:energies}). An excess energy on the order of $10^{23}$ -- $10^{24}\J$ is assumed to be sufficient for powering C-class flaring, which was actually observed for the active region analyzed here on the days before. Similar values related to C-class flaring activity were found by, \eg, \cite{reg_pri_07b,tha_wie_08b} and lately by \cite{gil_whe_12}.

\begin{table}
	\caption{\label{tab:energies} Magnetic energies of the 3D force-free models}
	\vspace{-11pt}
	\begin{center}
		\begin{tabular}{lcccc}
			\tableline
			~ & $\Enlff$ & $\Epot$ & $\Eexcess$ & $\frac{\Eexcess}{\Enlff}$\\ 
			~ & \multicolumn{3}{c}{$[\,\times$\,$10^{25}$\,$\J\,]$} &\\
			\tableline
			HMI& 3.45 & 2.80 & 0.65 & 0.19\\	
			SP$_{\rm bin}$ & 7.44 & 5.80 & 1.64 & 0.22\\ 
			SP$_{\rm orig}$ & 7.31 & 5.85 & 1.46 & 0.20 \\	
			\tableline
		\end{tabular}
	\end{center}
	{Given are the total, potential and excess magnetic energy of the 3D magnetic model fields, listed as $\Enlff$, $\Epot$ and $\Eexcess$, respectively. The ratio $\Eexcess/\Enlff$ gives the relative amount of excess energy. The statistical error of these estimates is $\sim$\,1\% for $\Epot$ and $\Enlff$ and $\sim$\,10\% for $\Eexcess$.}
\end{table}

\subsection{Magnetic Field Topology}\label{ss:topo}

\begin{figure*}
	\epsscale{1.0} 
	\plotone{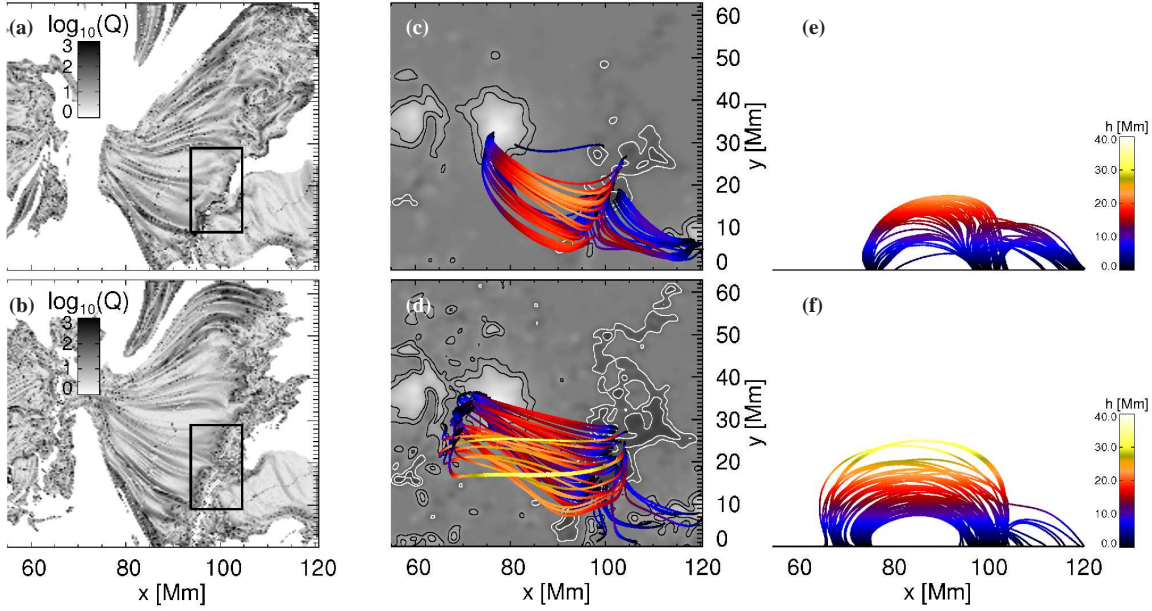}
	\put(-435,220){\bf(a)}
	\put(-285,220){\bf\white(c)}
	\put(-135,220){\bf(e)}
	\put(-435,115){\bf(b)}
	\put(-285,115){\bf\white(d)}
	\put(-135,115){\bf(f)}
	\caption{Squashing degree, $Q$, at a height of $\sim$\,$5\Mm$ above the NLFF lower boundary as determined from the (a) HMI and (b) SP$_{\rm bin}$ model, shown in the range $(55\Mm$\,$<$\,$x,y)$. Field lines originating from locations where $Q$\,$>$\,100 within the ROI (rectangular outlined area) and which have both footpoints located on the lower boundary of the HMI/SP$_{\rm bin}$ model are shown in (c,e)/(d,f). The magnetic configuration when viewed along the surface normal is shown in (c) and (d). The gray-scale background reflects the vertical magnetic field, $\Bz$, of the NLFF lower boundary. Black and white contours are drawn at [$+$50$\mT$,$+$100$\mT$] and [$-$50$\mT$,$-$100$\mT$], respectively. The view from solar south (along $y$) is shown in (e) and (f). Color-coded is the absolute height of the calculated field lines in$\Mm$.}
	\label{fig:fig4}
\end{figure*}

Besides comparing a volume-integrated quantity like the magnetic energy, we are also interested how the modeled magnetic field configurations compare to each other. To do so and to ensure that we are looking at the same topological structure we look for regions of strong gradients in the magnetic connectivity. They are thought of being linked to the creation of strong current concentrations in the solar corona and believed to represent the footprint of quasi-separatrix layers \citep{pri_dem_95}. We quantify the magnetic connectivity following \citet{tit_hor_02} and calculate the squashing degree $Q$ at a height of $\sim$\,5$\Mm$ above the NLFF lower boundary (Figure~\ref{fig:fig4}a,b). The squashing degree quantifies the eccentricity of an elliptical cross-section of a flux tube into which a flux tube of initially circular cross-section is transformed. Wherever $Q$ is large, the magnetic field connectivity changes drastically over short distances. According to this pattern, we choose a region of interest (ROI) around a clearly distinguishable pattern of high values of $Q$ (rectangular outline in Figure~\ref{fig:fig4}a,b). Though clearly visible in both models, the $Q$-ridge appears more diffuse in the SP$_{\rm bin}$ model and its location differs up to $\Delta x$\,$\simeq$\,$5\Mm$ for a given $y$.

\begin{figure}
	\centering
	\includegraphics[width=0.6\columnwidth]{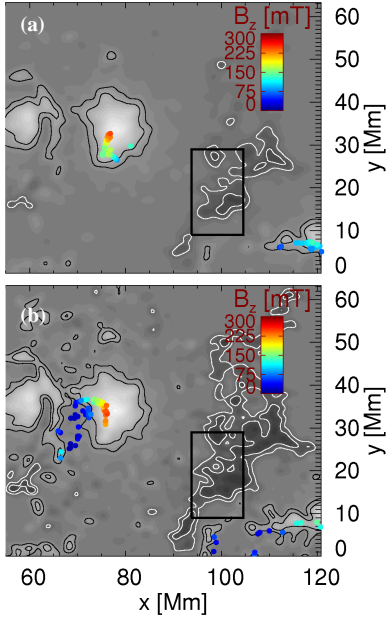}
	\put(-140,225){\bf\white(a)}
	\put(-140,115){\bf\white(b)}
	\caption{NLFF lower boundary of the (a) HMI and (b) SP$_{\rm bin}$ model in the range $(55\Mm$\,$<$\,$x,y)$. The vertical magnetic field is shown as gray-scale background. Black and white contours are drawn at [$+$50$\mT$,$+$100$\mT$] and [$-$50$\mT$,$-$100$\mT$], respectively. The locations where field lines originate from regions of $Q$\,$>$\,100 within the ROI (rectangular outline) re-enter the lower boundary are color-coded according to the vertical magnetic field there.}
	\label{fig:fig5}
\end{figure}

\begin{figure}
	\centering
	\includegraphics[width=\columnwidth]{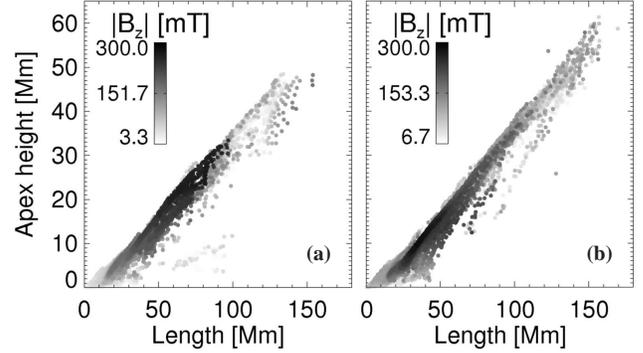}
	\put(-130,35){\bf(a)}
	\put(-24,35){\bf(b)}
	\caption{Relationship of the length of all closed model field lines, $\lfl$, to their apex height ($\hmax$) for the (a) HMI and (b) SP$_{\rm bin}$ model. Field lines are calculated starting from every pixel location in the range $(60\Mm$\,$<$\,$x,y)$ where $\Bzabs$\,$>$\,2$\sigma_{\Bzabs}$ {\it and} $\Bhabs$,$>$\,2$\sigma_{\Bhabs}$ or where $\Bzabs$\,$>$\,2$\sigma_{\Bzabs}$ {\it and} $\Bhabs$\,$<$\,2$\sigma_{\Bhabs}$ and the color-code reflects the value of the absolute vertical field there.}
	\label{fig:fig6}
\end{figure}

\begin{figure*}
	\epsscale{0.65} 
	\plotone{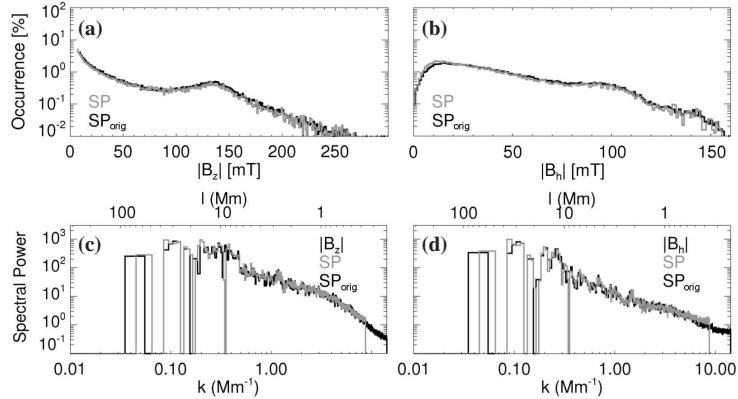}
	\put(-255,140){\bf(a)}
	\put(-125,140){\bf(b)}
	\put(-255,58){\bf(c)}
	\put(-125,58){\bf(d)}
	\caption{Relative occurrence of the absolute (a) vertical field $\Bzabs$, and (b) horizontal field, $\Bhabs$. The power of $\Bzabs$ and $\Bhabs$ as a function of length scale, $l$, and wave number, $k$, is shown in panels (c) and (d), respectively. SP$_{\rm bin}$ data and original-resolution data (SP$_{\rm orig}$) are represented by gray and black lines, respectively. Only pixels with values $\Bzabs$\,$>$\,2$\sigma_{\Bzabs}$ {\it and} $\Bhabs$,$>$\,2$\sigma_{\Bhabs}$ or $\Bzabs$\,$>$\,2$\sigma_{\Bzabs}$ {\it and} $\Bhabs$\,$<$\,2$\sigma_{\Bhabs}$ are taken into account.}
	\label{fig:fig7}
\end{figure*}

In total, 92/83 magnetic field lines in the HMI/SP$_{\rm bin}$ model ({\sc{i}}) start from locations where $Q$\,$>$\,100 within the ROI and ({\sc{ii}}) connect back to the lower boundary (see Figure~\ref{fig:fig4}c-f). They qualitatively outline two neighboring connectivity domains which connect the negative polarity region to its neighboring positive polarity surrounding. The locations where the field lines connect back to the lower boundary are shown color-coded based on the local vertical field in Figure~\ref{fig:fig5}. We assume that the field lines we calculated represent thin flux tubes. For each flux tube, we choose its cross section at the location from where we started the field line calculation, $\da$, as the size of one pixel (\ie, $\da$\,$\sim$\,360$^2$\,$\km^2$). Furthermore, we assume the vertical field there determines the flux of the flux tube. Summation over the all considered thin flux tubes gives an estimate of the total absolute shared flux for the HMI/SP$_{\rm bin}$ model, where we find $\fuvp$\,$=$\,$1.0$/$1.3$\,$\times$\,$10^{20}\Mx$. We find a lower value of connected flux in the HMI model though more closed magnetic field lines are considered due to our selection criterion. However, comparable is the {\it relative} amount of connected flux linked by these field lines: $\fuvp$ comprises $\sim$\,1\% of the unsigned vertical NLFF lower boundary flux as listed in Table~\ref{tab:forces}.

We also recognize differences in the connectivity pattern of the field lines as shown in Figure~\ref{fig:fig5}. Numerous SP$_{\rm bin}$ model field lines connect to the weak-field regions ($\Bzabs$\,$\lesssim$\,$50\mT$) between the two major positive polarity patches centered around ($x$\,$\sim$\,$65\Mm$, $y$\,$\sim$\,$35\Mm$) or to the weak-field surrounding at ($90\Mm$\,$\lesssim$\,$x$, $10\Mm$\,$\lesssim$\,$y$). However, none of the selected field lines of the HMI model does so, instead they re-enter the NLFF lower boundary at locations of strong vertical fields ($\Bzabs$\,$\gtrsim$\,$50\mT$). From Figure~\ref{fig:fig4}c-f one recognizes that a larger number of field lines connects to the positive polarity at ($x$,$y$)\,$\sim$\,($120\Mm$, $5\Mm$) and the highest field lines of the SP$_{\rm bin}$ model reach up to greater heights than those in the HMI model. 

To investigate if the latter represents a general trend or is biased due to our restrictive selection of start locations for field line calculation, we consider all field lines starting from a pixel location on the NLFF bottom boundary in the range $(60\Mm$\,$<$\,$x,y)$ where $\Bzabs$\,$>$\,2$\sigma_{\Bzabs}$ {\it and} $\Bhabs$,$>$\,2$\sigma_{\Bhabs}$ or $\Bzabs$\,$>$\,2$\sigma_{\Bzabs}$ {\it and} $\Bhabs$\,$<$\,2$\sigma_{\Bhabs}$. We then find very similar relationships of the length of the calculated field lines, $\lfl$, to the height of their apex, $\hmax$ for both models (see Figure~\ref{fig:fig6}): the relation is well defined and to a first approximation linear \citep[see also, \eg][]{schr_asch_02}. The SP$_{\rm bin}$ model field lines follow a steeper distribution, \ie, seem to be on overall higher. It appears that the field lines carrying the strongest vertical fluxes are neither the shortest ones nor the longest ones. Instead, it seems that in both models, field lines with a length of 20\,$\lesssim$\,$\lfl$\,$\lesssim$\,100$\Mm$ and an apex height of 5\,$\lesssim$\,$\hmax$\,$\lesssim$\,35$\Mm$ carry most magnetic flux. 

The longest and highest closed field lines in the HMI model are found to be $\sim$\,$150\Mm$ and $50\Mm$, respectively. This is lower than the values found for the longest/highest closed field lines of the SP$_{\rm bin}$ model ($\sim$\,$170\Mm$/$60\Mm$). It is not surprising that the SP$_{\rm bin}$ model field lines tend to be longer and reaching higher up in the model atmosphere when looking at the magnetic field distribution on the lower boundary. There, we find an average of $\Brave$\,$=$\,0.8/0.9 for the HMI/SP$_{\rm bin}$ model, \ie, the SP$_{\rm bin}$ model field lines are on average more vertical.

\subsection{Effects of SP data binning on the modeling}\label{ss:eof_binning}

\begin{figure*}
	\epsscale{1.0} 
	\plotone{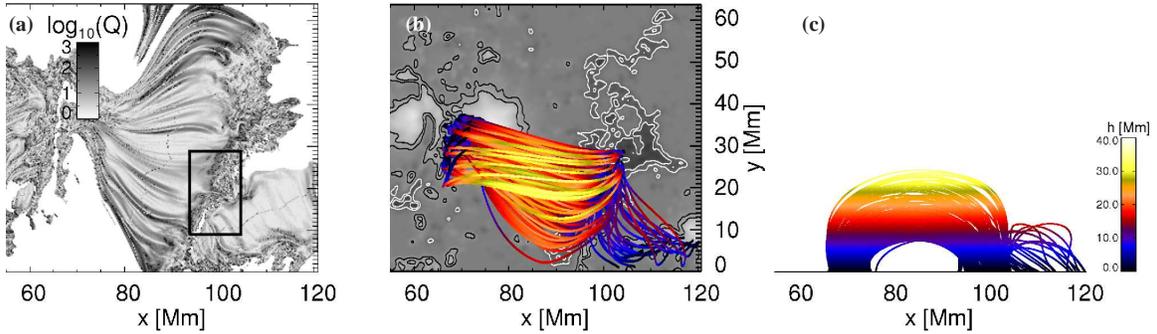}
	\put(-435,115){\bf(a)}
	\put(-285,115){\bf\white(b)}
	\put(-135,115){\bf(c)}
	\caption{(a) Squashing degree, $Q$, at a height of $\sim$\,$5\Mm$ above the NLFF lower boundary as determined from the closed SP$_{\rm orig}$ model field lines, shown in the range $x$\,$\ge$\,$55\Mm$. Field lines originating from locations where $Q$\,$>$\,100 within the ROI (outlined rectangular area) and which have both footpoints located on the lower boundary are shown in panels (b) and (c). The magnetic configuration when viewed along the surface normal is shown in (c). The gray-scale background reflects the vertical magnetic field, $\Bz$, of the NLFF lower boundary. Black and white contours are drawn at [$+$50$\mT$,$+$100$\mT$] and [$-$50$\mT$,$-$100$\mT$], respectively. The view from solar south (along $y$) is shown in (c).}
	\label{fig:fig8}
\end{figure*}

\begin{figure}
	\centering
	\includegraphics[width=0.59\columnwidth]{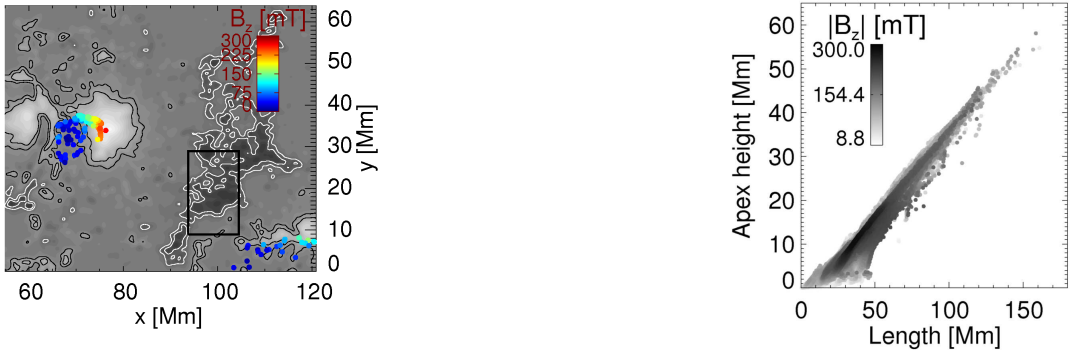}
	\caption{NLFF lower boundary of the SP$_{\rm orig}$ model in the range $(55\Mm$\,$<$\,$x,y)$. The vertical magnetic field is shown as gray-scale background. Black and white contours are drawn at [$+$50$\mT$,$+$100$\mT$] and [$-$50$\mT$,$-$100$\mT$], respectively. The locations where field lines originate from regions of $Q$\,$>$\,100 within the ROI (rectangular outline) re-enter the lower boundary are color-coded according to the vertical magnetic field there.}
	\label{fig:fig9}
\end{figure}

The binning of the original-resolution SP data (SP$_{\rm orig}$) to the resolution of HMI involves a 2D interpolation and the resulting changes are discussed in the following. As before, we only consider data values fulfilling the criteria outlined in \S~\ref{ss:uncertainty_estimation}.

The binning causes a decrease of $\fuv$ by $\sim$\,3\% (see Table~\ref{tab:forces}). It also causes an increase of $\BhaveSPorig$ by $\sim$\,6\%. The force-freeness remains basically the same, as do the relative occurrence of the vertical/horizontal field (Figure~\ref{fig:fig7}a/b) and the respective power distributions (Figure~\ref{fig:fig7}c/d) remain similar. Now it is also evident that the binning is only to a minor degree responsible for the differences between the SP$_{\rm bin}$ and HMI data. This agrees with, \eg, \cite{wan_zha_09a} who found that the scaling the data to a lower resolution does not significantly alter the results of their particular analysis.

The HMI data hosts $\sim$\,53\% of $\fuv_{\rm SP_{\rm orig}}$ and $\sim$\,86\% of $\BhaveSPorig$ (see Table~\ref{tab:forces}). Naively, one would suspect that the difference in $\fuv$ arises from the different resolution limits of the two instruments, \ie, that the magnetic field is partially on scales which HMI cannot resolve. In context with the relative occurrence of $\fuv$ discussed in \cite{tha_pie_12}, however, this cannot be concluded since it was found that HMI vertical fields are on overall weaker than those of VSM data (which with a plate-scale of $\sim$\,1~arc-second has a {\it lower} resolution than HMI). Therefore, the different occurrence rates must have reasons besides the different resolution limit of the instruments.

Preprocessing the SP$_{\rm orig}$ data leads to a decrease of $\fuv_{\rm SP_{\rm orig}}$ by $\sim$\,3\% and and increase of $\BhaveSPorig$ by $\sim$\,40\%, which is comparable to the changes of the SP$_{\rm bin}$ data due to preprocessing. The modification to the SP$_{\rm orig}$ lower boundary data during solving for the force- and divergence-free field, as listed in Table~\ref{tab:forces}, is that $\fuv_{\rm SP_{\rm orig}}$ decreases by $\sim$\,3\% and $\BhaveSPorig$ increases by $\lesssim$\,1\%. Summarizing, we find a similar behavior of the modifications to the SP$_{\rm orig}$ data during the force-free modeling and very similar values as we found for the SP$_{\rm bin}$ data (compare the values listed in Table~\ref{tab:forces}).

The potential energy of the SP$_{\rm bin}$ model is $\sim$\,1\% lower and the total energy is $\sim$\,2\% higher than that of the SP$_{\rm orig}$ model (see Table~\ref{tab:energies}). Hence, when basing the analysis on the SP$_{\rm bin}$ data, we find and enhancement/reduction of the absolute energy estimates $\Enlff$/$\Epot$ on the order of the statistical error of the energy estimates itself. The excess energy, $\Eexcess$, is enhanced by $\sim$\,10\%, \ie, also on the order of the statistical error. The {\it relative} amount of $\Eexcess$, however, remains approximately the same.

\begin{figure}
	\centering
	\includegraphics[width=0.56\columnwidth]{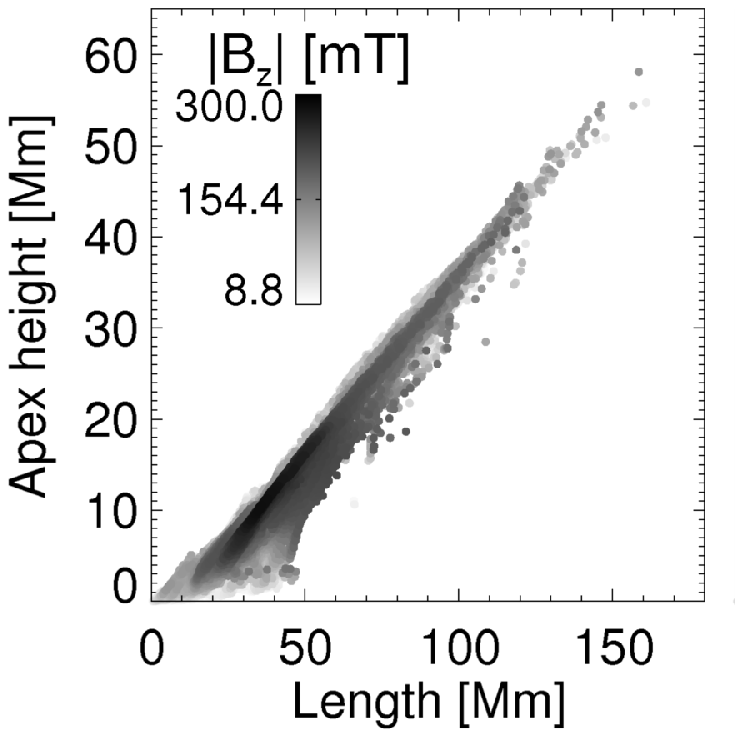}
	\caption{Relationship of the length of closed model field lines, $\lfl$, to their apex height ($\hmax$) for the SP$_{\rm orig}$ model. Field lines are calculated starting from every pixel location in the range $(60\Mm$\,$<$\,$x,y)$ where $\Bzabs$\,$>$\,2$\sigma_{\Bzabs}$ {\it and} $\Bhabs$,$>$\,2$\sigma_{\Bhabs}$ or where $\Bzabs$\,$>$\,2$\sigma_{\Bzabs}$ {\it and} $\Bhabs$\,$<$\,2$\sigma_{\Bhabs}$. The color-code reflects the value of the absolute vertical field there.}
	\label{fig:fig10}
\end{figure}

Repeating the analysis of the magnetic connectivity within the SP$_{\rm orig}$ model, we find that in total 228 magnetic field lines originating from locations of $Q$\,$>$\,100 within the ROI in Figure~\ref{fig:fig8}a connect back to the lower boundary (see Figure~\ref{fig:fig8}b,c). Again, we assume that those represent thin flux tubes with a cross section of $\da$\,$\sim$\,220$^2$\,$\km^2$ and assume the vertical field at footpoint from which we started the field line calculation determines its flux. Summation over the all considered thin flux tubes gives an estimated total shared flux of $\fuvp$\,$=$\,$1.2$\,$\times$\,$10^{20}\Mx$ which comprises about 1\% of the unsigned vertical flux of the NLFF lower boundary of the SP$_{\rm orig}$ model. This is almost identical to what was found for the model based on the SP$_{\rm bin}$ data, as is the connectivity pattern (compare Figures\,\ref{fig:fig9} and \ref{fig:fig5}b). Considering all field lines starting from a pixel location on the NLFF bottom boundary in the range $(60\Mm$\,$<$\,$x,y)$ where $\Bzabs$\,$>$\,2$\sigma_{\Bzabs}$ {\it and} $\Bhabs$,$>$\,2$\sigma_{\Bhabs}$ or $\Bzabs$\,$>$\,2$\sigma_{\Bzabs}$ {\it and} $\Bhabs$\,$<$\,2$\sigma_{\Bhabs}$, we find an identical relationship of $\lfl$and $\hmax$ (Figure~\ref{fig:fig10}) as was found for the SP$_{\rm bin}$ model (compare Figure~\ref{fig:fig6}b). The SP$_{\rm orig}$ model field is found to be even more vertical ($\Brave$\,$\sim$\,0.97) and, again, the field lines carrying the strongest vertical fluxes are those with an intermediate length and apex height.

\section{Discussion and Conclusions}

Non-identical photospheric vector magnetic fields inferred from polarization measurements of different instruments used as input for nonlinear force-free (NLFF) coronal magnetic field models directly translate to substantial differences in the model outcome. To quantify these, we performed force-free magnetic field modeling using active-region vector magnetic field data based on measurements of polarization signals by the SDO/HMI and Hinode SOT/SP. The possible causes of the differences of the data products itself, including the, \eg, different intrinsic nature and sensitivity of the instruments, the temporal evolution of the photospheric magnetic field during the ongoing scanning times or the inversion techniques used to infer the magnetic field vector from the measured polarization signals, were out of the scope of this work. Our aim was to compare force-free model results based on data from these two instruments so we applied all data preparation and modeling routines in exactly the same way to both data sets. 

Force-free coronal magnetic field modeling is computationally expensive and high-resolution data is sometimes binned to a lower resolution in order to shorten the computational time. We, therefore, binned the original-resolution SP (SP$_{\rm orig}$) data to the resolution of HMI which allowed us, besides ({\sc{i}}) to quantify the deviations of the force-free modeling outcome due to the usage of data of {\it different} instruments (HMI and SP), also ({\sc{ii}}) to investigate the effect of binning the input data to a lower resolution on the model outcome, by subsequent comparison of the models based on the binned SP (SP$_{\rm bin}$) data and SP$_{\rm orig}$ data. We did not intend to mimic the different spatial resolution of the instruments by binning of the vector data since, as pointed out by \cite{lek_bar_12}, any kind of binning does not account properly for resolution effects. Instead, in practice, a binning of high-resolution data prior to the NLFF modeling is meant to result in feasible computational dimensions and to allow for near real-time modeling.

We used HMI and SP vector maps of active region~11382 on 2011 December 22 and found considerably higher vertical magnetic flux and average horizontal field in the SP data. This difference of the vertical flux was found to be much larger than the modifications to the data due to application of our force-free modeling routines. Also the modifications to the vertical flux due to binning of the SP$_{\rm orig}$ data to the resolution of HMI were found to be small compared to the inequality of detected flux by the two instruments. 

Unequal estimates of the magnitude and orientation of longitudinal and transverse fields based on the inversion of circular and linear polarization signals measured by the two different instruments, yield differing estimates of the vertical and horizontal magnetic field in a local coordinate system. The unequal amount of vertical magnetic flux (the projected SP data hosting about two times the unsigned vertical flux of HMI) then directly translates to a considerable discrepancy in the {\it absolute} estimates of the energy content of the considered coronal volume: models based on the SP data hold about twice as much energy as does the HMI model. By tracing magnetic field lines in the half-space above the model lower boundaries their connectivity was investigated. 

The SP and SP$_{\rm orig}$ models revealed a closely matching photospheric footprint of two neighboring coronal connectivity domains. A similar footprint was found in the HMI model but locally shifted by up to several$\Mm$. The same applies to the locations where field lines re-enter the NLFF lower boundary: while the location of the footpoints of the SP and SP$_{\rm orig}$ models coincide, those of the HMI model are displaced by up to ten$\Mm$. Moreover, on overall, the SP and SP$_{\rm orig}$ model fields tend to be more vertical than the HMI model field which is a direct consequence of the relative vertical and horizontal field distribution given by the instrument data.

However, the models also showed great similarities: {\it relative} estimates like the fraction of energy in excess over a potential field (about 20\% of the total energy content) or the fraction of vertical flux shared by two neighboring connectivity domains (about 1\% of the total vertical flux of the active region) agree very well. Also the overall field-line geometry was found to be comparable: the length of all closed field lines in the model volumes was found to be more or less linearly related to the apex height. Common to the model outcomes is also that the shortest and lowest as well as the longest and highest field lines carry least vertical flux.

In conclusion, caution is needed when analyzing the coronal magnetic field and its connectivity with the help of force-free magnetic field models based on the vector magnetic field products of different instruments, made available to the community. {\it Relative} estimates and the overall structure of the model magnetic fields might indeed be reliable while {\it absolute} estimates might only be so concerning their order of magnitude. Moreover, binning of the magnetic vector data to a lower resolution prior to the force-free modeling results only in little differences in the model outcome, small compared to the remarkable deviations when basing the modeling on data from the two different instruments, HMI and SP.

\acknowledgments
We thank the anonymous referee for careful consideration in order to improve our manuscript. We would also like to thank X.~Sun for support with the HMI vector magnetic field data and B.~Inhester for insightful discussions.  J.\,K.\,T.~acknowledges supported by DFG grant WI 3211/2-1, T.\,W.~is funded by DLR grant 50 OC 0904. SDO data are courtesy of the NASA/SDO HMI science team. Hinode is a Japanese mission developed and launched by ISAS/JAXA, with NAOJ as domestic partner and NASA and STFC (UK) as international partners. It is operated by these agencies in co-operation with ESA and NSC (Norway). Hinode SP Inversions were conducted at NCAR under the framework of the CSAC.\\

\bibliographystyle{apj}

\end{document}